# Defects responsible for abnormal *n*-type conductivity in Ag-excess doped PbTe thermoelectrics

Byungki Ryu (류병기),[1,a)] Min-Wook Oh (오민욱),[2] Jae Ki Lee (이재기),[1] Ji Eun Lee (이지은),[1] Sung-Jae Joo (주성재),[1] Bong-Seo Kim (김봉서),[1] Bog-Ki Min (민복기),[1] Hee-Woong Lee (이희웅), [1] and Su-Dong Park (박수동)[1]

[1]*Thermoelectric Conversion Research Center, Korea Electrotechnology Research Institute (KERI), 642-120, Changwon, Republic of Korea*

[2]*Department of Advanced Materials Engineering, Hanbat National University, 305-719, Daejeon, Republic of Korea*

We find that Ag-interstitial ($Ag_I$) acts as an electron donor and plays an important role in Ag-excess doped polycrystalline PbTe thermoelectric materials. When Ag is heavily doped in PbTe, the neutral (Ag-Ag) dimer defect is formed at the Pb-site and the environment becomes Pb-rich/Te-poor condition. Then the positively ionized Ag interstitial ($Ag_I^+$) defect becomes the major defect under Pb-rich condition. Due to the small formation energy and small diffusion barrier of $Ag_I^+$, Ag can be easily dissolved into the PbTe matrix. The temperature behavior of the Ag defect formation energy well explains the $Ag_I^+$ solubility, the electron carrier generation, and the increasing electrical conductivity in Ag-excess doped polycrystalline PbTe at high temperature. This abnormal doping behavior by forming interstitial defects is also found for Au-doped PbTe.

a) Corresponding author email: byungkiryu@keri.re.kr or cta4rbk@gmail.com.

Thermoelectric effect refers to the direct energy conversion between heat and electricity. Its technology has attracted much attention because of its potential application to the power generation from waste heat.[1] [Snyder and Toberer (2008)] The efficiency of thermoelectric system is governed by the dimensionless figure of merit $ZT$, defined as $ZT = (\alpha^2\sigma/\kappa)T$ where $\alpha$, $\sigma$, $\kappa$, and $T$ are, Seebeck coefficient, electrical conductivity, thermal conductivity, and absolute temperature, respectively.[2][Goldsmid (2010)] There have been many strategies to increase the average or peak $ZT$. Optimization of power factor ($\alpha^2\sigma$) is possible by intrinsic and extrinsic doping.[3][Oh (2014)][4,5][Lee (2014), Pei (2011c)] Enhancement of Seebeck coefficient is achievable by quantum confinement effect,[6][Hicks (1993)] formation of resonant state,[7][Heremans (2008)] band structure convergence,[8,9][Pei (2011a), Liu (2012)] carrier energy filtering,[10][Faleev (2008)] and minority carrier blocking.[11][Yang (2015)] Reduction of lattice thermal conductivity is obtained by grain boundary scattering from nanostructuring[12,13][Hsu (2004), Wang (2014)] and all-scale hierarchical architecture.[14][Biswas (2012)] Most of above approaches are based on the atomic structure modification by defect formation in thermoelectric materials.

The Ag is a good doping/alloying elements in telluride thermoelectric materials because Ag-related or Ag-rich impurity phases such as $AgSbTe_2$ and $Ag_2Te$ can form a coherent interface with PbTe and GeTe and thereby the Ag-related defect can selectively scatter the phonons, while the reduction of electronic transport is relatively small.[12,15,16][Hsu (2004), Cook (2007), Pei (2011b)]

Ag is also known as a dopant in tellurides. Some studies have shown that Ag doped PbTe exhibits *p*-type characteristics as Ag atoms substitutes cation atoms,[17][Dow (2010)] similar to Na doped PbTe.[5,13],[Pei (2011c), Wang (2014)] However, the doping of Ag is ineffective and the carrier concentration is below $10^{18}$ cm$^{-3}$.[17][Dow (2010)] Ag also shows *n*-type behavior in PbTe. For example, Ag-excess doping in PbTe leads the formation of *n*-type Ag defects as reported Pei et al.[18]. The most surprising result is that the electron chemical potential is self-tuned by Ag doping in PbTe/$Ag_2Te$ composite and the electrical conductivity increases when temperature increases. Note that most of thermoelectric materials are degenerate semiconductor and their conductivities decrease when temperature increases due to the electron-phonon interaction. Pei *et al*. explained that it is due to the formation of Ag interstitial (Ag$_I$) donor at $T$ = 450–750 K, rather than formation or activation of native defects.[18] However there is an inconsistency between the fitted energy slope of Ag$_I$ formation ($E_{slope}$ = 0.23 eV, N$_e$~5×10$^{20}$ cm$^{-3}$ at T = 700 K) and the electron carrier concentration (~2×10$^{19}$ cm$^{-3}$ at T = 700 K). Unfortunately, there has been lack of understandings for *n*-type Ag dopant in PbTe. A few theoretical and computational studies have been performed to understand the electrical properties of Ag and Ag-related defects in PbTe, however these works were done for resonant defects, deep defects, or impurity phase defects consisted of Ag-substitutionals.[19,20,21,22][Bilc (2004), Ahmad (2006), Hoang (2007), Hoang (2010)] In our knowledge, the defect properties of Ag interstitials in PbTe are not studied yet.

In this Letter, we report the results of first principles calculations for the Ag defect in PbTe. We find that the Ag interstitial is the effective donor in Ag-excess doped in PbTe. The formation energy and migration energy barrier calculations reveal that the ionized Ag interstitial is generated at high temperature and can reach whole region of polycrystalline PbTe. Finally, we elucidate that the formation of Ag interstitial well explains the abnormal *n*-type carrier concentration at temperature between 400 and 800 K, which is responsible for Fermi-level self-tuning in Ag-excess doped PbTe thermoelectrics.

We perform first principles calculations based on the density functional theory.[23][Hohenberg and Kohn (1964), Kohn and Sham (1965)] We use the projector-augmented-wave (PAW) pseudopotential,[24][Blöchl (1994)] generalized-gradient-approximation parameterized by Perdew-Burke-Ernzerhof (PBE),[25][Perdew (1996)] which are implemented in VASP planewave code.[26][Kresse (1996), Kresse (1999)] Defective structures based on 64-atom cubic PbTe supercell are used with the kinetic energy cutoff of 400 eV and 4×4×4 MP k-point mesh.[27][Monkhorst and Pack (1976)] The PbTe lattice parameter is fixed to 6.575 Å and the atomic positions are fully relaxed until the atomic forces become smaller than 0.005 eV/Å . Electronic structures and formation energies are calculated with inclusion of spin-orbit-interaction. The PbTe band gap ($E_g$) is corrected to 0.397 eV, using HSE06 hybrid density functional calculations.[28][Krukau (2006)] The formation energy ($E_{form}$)of charged defect ($D^q$) is calculated using following equations[29][Zhang (1991)]:

$$E_{form}[D^q] = E[D^q] - E_O - \Sigma_i \Delta n_i \mu_i + q(E_F + E_{VBM} + \delta V), \tag{1}$$

where $E[D^q]$ and $E_O$ are the total energy with and without defect $D^q$, $i$ is the index of atomic species, $\Delta n_i$ is number change of $i$ in the formation of defect in defective supercell with respect to ideal bulk, $\mu_i$ is atomic chemical potential, $q$ is the defect charge state, $E_F$ is the Fermi level, and $E_{VBM}$ is the energy of valence band maximum (VBM). At this moment, the sum of atomic chemical potential of Pb and Te is constrained to the heat of formation of PbTe with respect to the metallic Pb and Te phases. And Ag chemical potential is set to the total energy of BCC Ag, to model to Ag-rich environment. To correct the charged supercell error in formation energy, we use the potential alignment of $\delta V$ from local potential shift.[30][Komsa (2012)] For 1+ charge defect $Ag_I^+$, the energy difference between corrected formation energies of 64 and 512 atom supercells is less than 0.01 eV. Nudged elastic band (NEB) method is adopted to calculate the energy barrier of defect migration ($E_m$).[31][Henkelman and Jónsson (2000a, 2000b)] The diffusion constant $D$ is calculated as

$$D = a^2 C \Gamma \tag{2}$$

where $a$ is the lattice constant and $C$ is a defect concentration determined from the interstitial site density and the defect formation energy.[32][Mantina (2008)] The transition rate $\Gamma$ is calculated from the vibration frequency of defect at the ground and saddle point, and the defect migration energy $E_m$. At this moment, the effective frequency $v^*$ is obtained from the density functional perturbation theory.[33][Gonze and Lee (1997)] The solubility of the defect is calculated based on the temperature and chemical electron potential dependent formation energy and Boltzmann distribution. The electron chemical potential for given electron carrier concentration is determined using Fermi-Dirac statistics and the density of states.

**Ag interstitial is an *n*-type dopant**. We investigate the electronic structures of Ag point defects in PbTe and find that Ag atoms can be either donors, acceptors, or isovalent defects. The electronic band structures of bulk PbTe, neutral $Ag_I^0$ in PbTe, and neutral Te-vacancy ($V_{Te}^0$) in PbTe are shown in **Figure 1(a-c)**. The band gap is calculated to be 0.1 eV for PBE exchange correlation energy [**Figure 1(a)**]. Due to the band folding, the band edges are located at the gamma point of ($2\times2\times2$) cubic supercell. When we adopt the hybrid DFT, the band gap is increased to 0.397 eV, which is comparable to the experimental band gap of 0.3 to 0.4 eV at T = 300 – 800 K. **Figure 1(b)** shows the shallow donor nature of the $Ag_I$ defect. Due to the electron transfer from $Ag_I$ to the PbTe conduction band minimum (CBM), the $E_F$ is shifted upward and located at within the conduction band. There is no Ag defect state near the band gap and at the energy ranging from $E_{CBM}$ to $E_{CBM}$+0.5 eV. Meanwhile, $Ag_{Pb}$ exhibits an acceptor behavior. As a monovalent-Ag substitutes a divalent-Pb in PbTe, $Ag_{Pb}^0$ generates one hole at the PbTe VBM. We also find that there is an attractive interaction between $Ag_{Pb}^-$ and $Ag_I^+$ and thereby the neutral isovalent (Ag-Ag) defect pair at Pb-site [(Ag-Ag)$_{Pb}$] can be formed. It is worth to note that the atomic structure of (Ag-Ag)$_{Pb}$ is very similar to the atomic structure of $Ag_2Te$ phase. **Figure 1(c)** shows the band structure of the $V_{Te}$ defect in PbTe. $V_{Te}$ is also a donor in PbTe. However, there are defect levels near the CBM, which can cause the Fermi level pining at highly *n*-doped PbTe. The position of Fermi level for $V_{Te}^0$ (0K electron chemical potential) is lower than that for $Ag_I^0$.

**Ag is dissolved as a form of neutral (Ag-Ag)$_{Pb}$ in Te-rich PbTe.** We calculate the $E_{form}$ of Ag defects [$Ag_I$, $Ag_{Pb}$, $Ag_{Te}$, (Ag-Ag)$_{Pb}$] with the charge state *q* from 2+ to 2− , as compared to the metallic Ag phase. The calculated $E_{form}$s are always positive. Under Te-rich condition, the formation of neutral (Ag-Ag)$_{Pb}$ defect is the most probable when $E_F$ is above the mid gap, explaining the $Ag_2Te$ dissolution in PbTe [**Figure 2(a)**].[34][Bergum (2011).] We predict that the solubility of (Ag–Ag)$_{Pb}$ defect to be 2 to 5 % at 600 K and 800 K, respectively. Note that, under Te-rich condition, the charge compensation is easily occurred between Ag defects in the Te-rich PbTe. When $E_F$ is at VBM, the hole carrier from $Ag_{Pb}^{1-}$ is easily compensated by the electron carrier from $Ag_I^+$. Similarly, when $E_F$ is at CBM, the electron carrier from $Ag_I^+$ is also easily compensated by the hole carrier from $Ag_{Pb}^-$. Thus, Ag is not an efficient acceptor in $Pb_{1-x}Te$ due to the charge compensation as well as the formation of the isovalent $(Ag-Ag)_{Pb}$ defect. Rather, it is thought that the origin of $p$-type conductivity is the formation of $p$-type Ag-impurity phases in $Pb_{1-x}Ag_xTe$ (x=0.1).[17][Dow (2010)]

**$Ag_I^+$ is the most stable defect in Pb-rich PbTe.** When $Ag_2Te$-like point defect or $Ag_2Te$ impurity phases are formed, the chemical potential condition will be changed to Pb-rich (Te-poor) condition. Under Pb-rich condition, the ionized $Ag_I^+$ becomes major [See **Figure. 2(b)**]. When $E_F$ is at $E_{CBM}(=E_{VBM}+0.397eV)$, the formation energies of $Ag_I^+$, $Ag_{Pb}^-$, $(Ag-Ag)_{Pb}^0$ are 0.519, 0.939, and 0.892 eV, respectively. The transition level $\varepsilon(q/Q)$ from $q$ to $Q$ is defined as the Fermi-level for which these $E_{form}$ are equal. For $Ag_I$ defect, $\varepsilon(2+/1+)$ and $\varepsilon(1+/0)$ are calculated to be $E_{VBM}-0.081$ and $E_{VBM}+0.511$ eV, respectively, meaning that $Ag_I$ is always behaving 1+ charged state as a shallow donor. Therefore, in Ag-excess doped $PbTe/Ag_2Te$ composites, $Ag_I$ is an effective $n$-type dopant even in degenerate limit, where Fermi level is well above the CBM.

**$Ag_I^+$ is a major donor in Pb-rich $n$-PbTe.** We calculate the formation energies of native defects in PbTe, including Pb- and Te-vacancies ($V_{Pb}$ and $V_{Te}$), Pb- and Te-interstitials ($Pb_I$ and $Te_I$), and Pb- and Te-antisite defects ($Pb_{Te}$ and $Te_{Pb}$) with charge state $q$ from 2+ to 2- [see **Figure 2(c)**]. In Te-rich condition, $Te_{Pb}^{2+}$ donor and $V_{Pb}^{2-}$ are the most stable among native defects in $p$-PbTe and $n$-PbTe, respectively. However, formation energies of these charge native defects are larger than the stable Ag defects. In Pb-rich PbTe, $V_{Te}^{2+}$ and $Ag_I^+$ are the most stable donor defects among native defects and Ag defects in $p$-PbTe and $n$-PbTe, respectively [see **Figure 2(d)**].

**Next we show that Ag easily dissolves into PbTe matrix.** We calculate the diffusion constant of $Ag_I^+$. Our nudged elastic band method (NEB) calculation reveals that the $Ag_I^+$ diffuses through the interstitialcy diffusion mechanism [see **Figure 3(a)**]. The ground state configuration is at the body center of $(PbTe)_4$ cube and the saddle point is at the face center of $(PbTe)_2$ square face. The migration energy barrier is calculated to be 0.558 eV for the lattice constant of 6.575 Å [see **Figure 3(b)**]. Thus when Fermi level is at the CBM, the diffusion energy barrier of $Ag_I^+$ is calculated to be 1.077 eV at 0 K. We also consider the effect of lattice parameter expansion, $a(T) - a_o = (0.9545 \times 10^{-4})T$, and calculate the temperature dependent migration barrier and defect formation energy. The $E_m$s are calculated to be 0.523, 0.501, and 0.484 eV at $T = 300, 500, 700$ K, respectively. By taking into account the real vibration frequencies of ground and saddle point configurations ($\nu_{g,i}$ and $\nu_{s,j}$, where $i = 1$ to 3N-3 and j = 1 to 3N-4) the effective frequency of $Ag_I^+$ diffusion ($\nu^* = \Pi_i\ \nu_{g,i}\ /\ \Pi_j\ \nu_{s,j} = 2.755$ THz) is obtained. Note that the effective lattice parameter for the Ag interstitial is a half of the lattice parameter because there are 8 interstitials in the conventional PbTe cubic cell. Finally, the diffusion constant (D = $a^2C\Gamma$) is evaluated using $a = a(T)/2$, the defect concentration $C = [2/a(T)]^3 \times \exp[-E_{form}(T)/k_BT]$, and the jump frequency $\Gamma = \nu^* \times \exp[-E_m(T)/k_BT]$. The diffusion coefficient of $Ag^{I+}$ is $3.37 \times 10^{-14}$ m$^2$/s at 700 K. Based on the random walk model [$l = (6Dt)^{1/2}$], we find that the diffusion length is comparable to the size of the grain size of polycrystalline PbTe. The diffusion length for 1 minute at 700 K is calculated to be 3.5 μm. In Ag-excess doped PbTe, there are metallic-Ag and Ag$2$Te impurity phases. Therefore, in Ag-excess doped PbTe, Ag can be easily dissolved in to the PbTe matrix from the impurity phase of metallic Ag and the $Ag_I^+$ shallow donor can be distributed over the whole lattice in a short time.

**Ag interstitial is generated at high temperature and responsible for abnormal doping nature.** Based on the electronic density of states and the Fermi-Dirac distribution of electron carrier, we calculate the carrier density as a function of electron chemical potential for given temperatures [see **Figure 4(a)**]. The electron chemical potential moves toward the mid gap when temperature increases due to the generation of minority carriers. We find that that the electron chemical potential and the charged defect formation energy can vary by about 0.1 eV due to the temperature and doping concentration, which is comparable to $4k_BT$ at room temperature. It means that the solubility is very sensitive to the electron chemical potential. To model the Ag defect abundant situation, we consider that there is only $Ag_I^+$ defect as a charged defect. To model the experimental situation, the *n*-type condition, we assume that the electron chemical potential is at the CBM or above the CBM. After taking into account the carrier density curve and the formation energy curve for given electron chemical potentials and given temperatures, we obtain the electron chemical potential and the self-consistent solubility of $Ag_I^+$ for temperature ranging from 400 to 700 K [see **Figure 4(b)**]. As shown in the Figure, our calculated solubility (solid line) well matches with the experimentally observed electron carrier concentration of Ag-excess doped PbTe. The red-dashed-line represents the solubility curve with the energy slope of 0.23 eV,[18][Pei (2011d)] which overestimates the electron carrier concentration by an order of 1 or 2. The blue-dotted-line stands for the formation energy when the electron chemical potential is at the CBM ($E_{form}$ = 0.519 eV), which underestimate the carrier concentration at the temperature below 700 K. In contrary, our calculated $E_{form}$[T] well describes the temperature dependent electron carrier concentration at the wide temperature range from 400 to 700 K.[18] [Pei (2011d)].

We would like to emphasize that the doping mechanism of Ag in PbTe is different from the dopant activation. In semiconductors such as Si, GaN, and ZnO, the dopant ionization energy plays a critical role. Thereby, the carrier concentration is increased with increasing temperature. However, in Ag-excess doped PbTe, the formation of Ag-defect determines the carrier concentration of PbTe materials. As the lattice size of PbTe is larger than other semiconductors, dopants can be mobile and dissolved in PbTe. We also study the formation and diffusion properties of Au and find that noble metals are very mobile and diffusible in PbTe. The formation energy of the $Au_I^+$ defect is 0.686 eV when $E_F$ is at CBM. The migration energy barrier is smaller by 0.1 to 0.2 eV for Au than for Ag. In Au-exceed doped PbTe, we expect that the $Au_I^+$ defects also tune the Fermi level in Au-excess doped PbTe, as $Ag_I^+$ defects do in Ag-excess doped PbTe.

In conclusion, based on the first-principles calculations, we reveal that the $Ag_I^+$ defect is the shallow donor and the major charged defect in *n*-type PbTe. When Ag is excess doped, $Ag_2Te$-like defect is formed. Then $Ag_I^+$ becomes major even when Fermi level is located at above the CBM. The diffusion length for a few minute is comparable to the grain size of polycrystalline-PbTe, indicating that Ag is easily dissolved into the PbTe. As the dissolved $Ag_I^+$ is distributed over the whole PbTe, the electron carrier concentration increases with increasing temperature.

**ACKNOWLEDGMENTS**

This work was supported by the Energy Efficiency & Resource Core Technology Program of the Korea Institute of Energy Technology Evaluation and Planning (KETEP) granted from the Ministry of Trade, Industry & Energy, Republic of Korea (no. 20112010100100). Also this work was supported by the National Institute of Supercomputing and Network (NISN) / Korea Institute of Science and Technology Information with supercomputing (KISTI) resources including technical support (KSC-2014-C1-022).

**FIGURES**

Fig.1. Band structures of (2×2×2) cubic supercell are drawn for (a) bulk PbTe, (b) $Ag_I^0$ in PbTe and (c) $V_{Te}$ in PbTe. The $E_F$ is denoted by horizontal-green-line and the VBM energy is set to zero.

Fig.2. Formation energies are drawn for various defects with various charge states: (a) Ag defects in Te-rich PbTe, (b) Ag defects in Pb-rich PbTe, (c) native defects in Te-rich PbTe, and (d) native defects in Pb-rich Pbte.

Fig.3. (a) Atomic configurations of $Ag_I^+$ during duffusion: the left one for ground state at initial state, the second one for saddle point, and the right one for final configuration. (b) The relative total energy of the $Ag_I^+$ defect during diffusion.

Fig.4 (a) Charge carrier densities are shown as a function of the electron Fermi level for given temperature raning from 300 to 800 K. (b) Density of $Ag_I^+$ and electron carriers are shown. The blue-solid-like with closed circle represents the solubility of $Ag_I^+$ calculated from this work. The black-solid-line with open circle represents the carrier concentration due to Ag-excess doping in PbTe, reported by Pei (2011d). The black-dashed-line with open circles represents the predicted carrier density from the energy slope of 0.23 eV. The blue-dashed-line with closed circles represents the predicted carrier density from the formation energy of $Ag_I^+$ when $E_F$ is at CBM (0.519 eV).

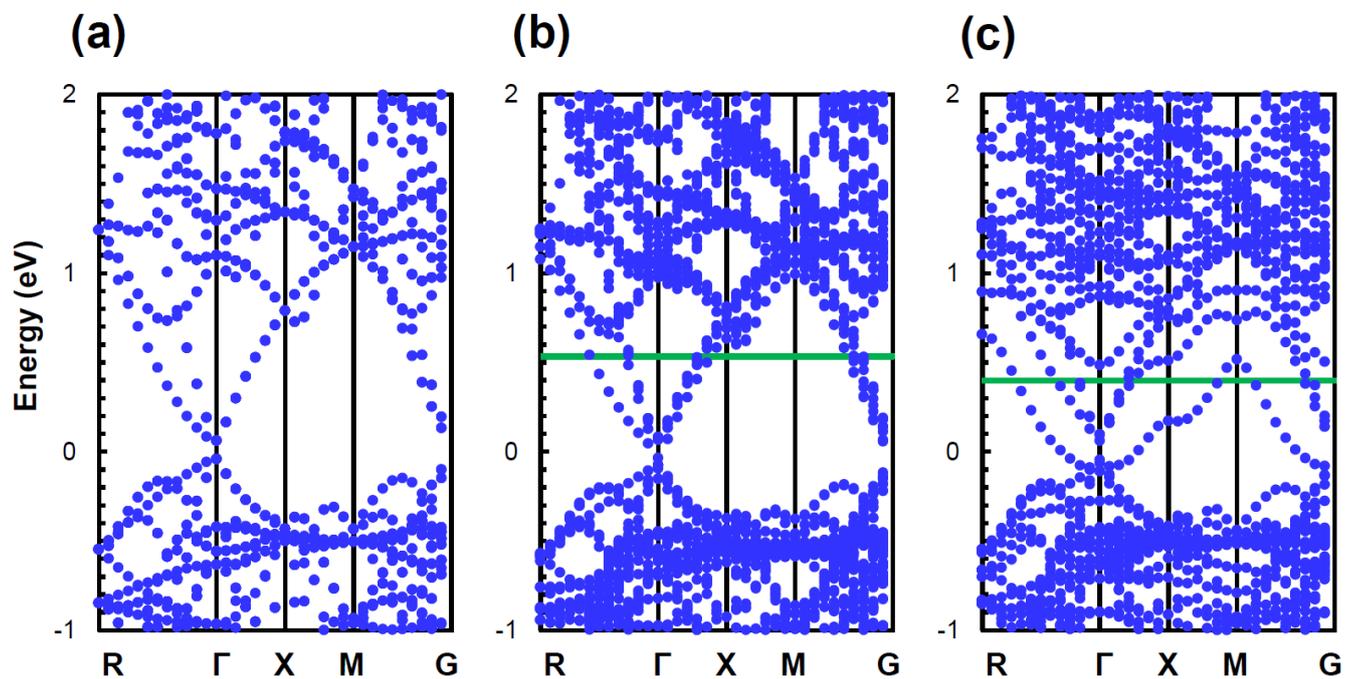

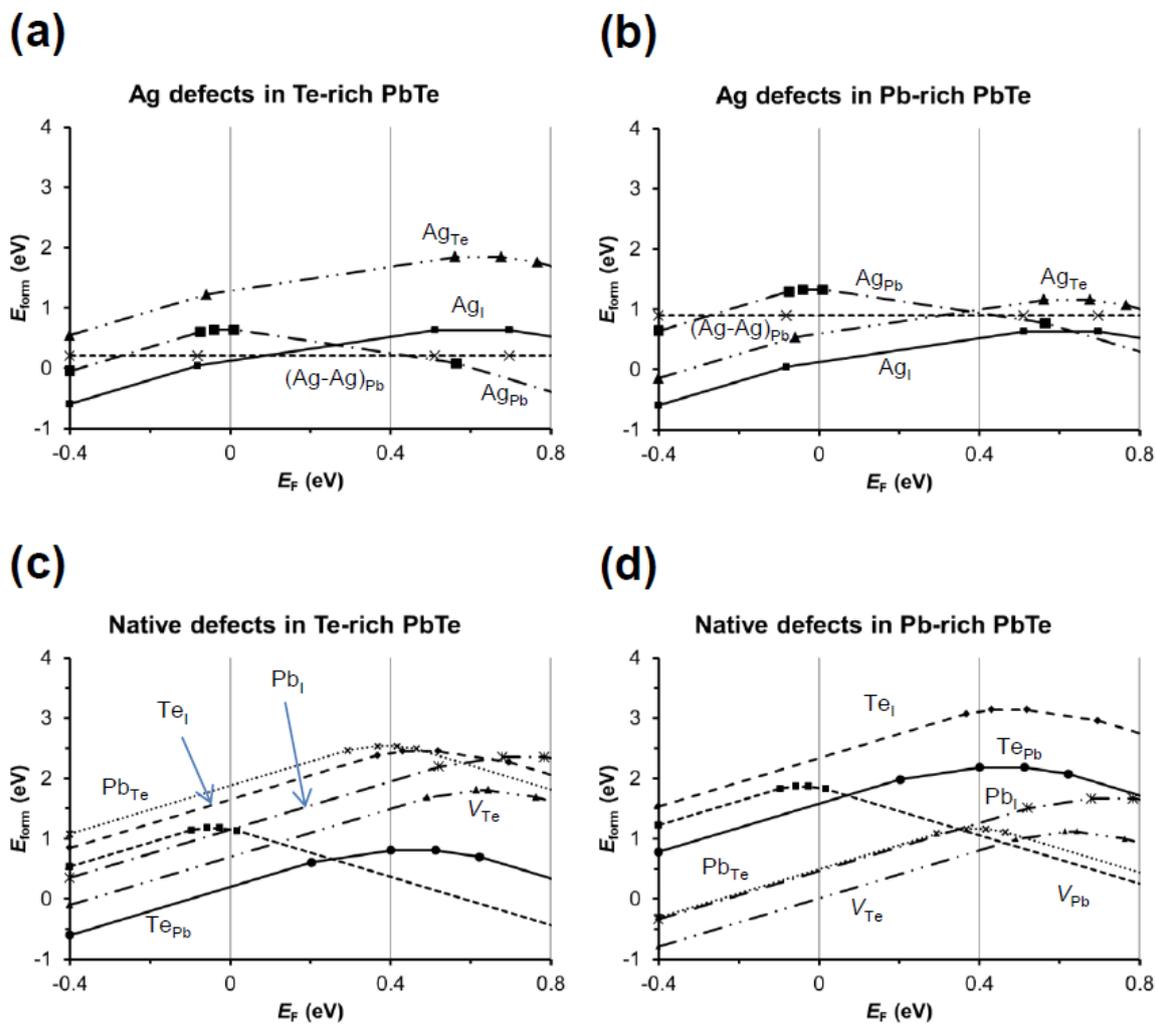

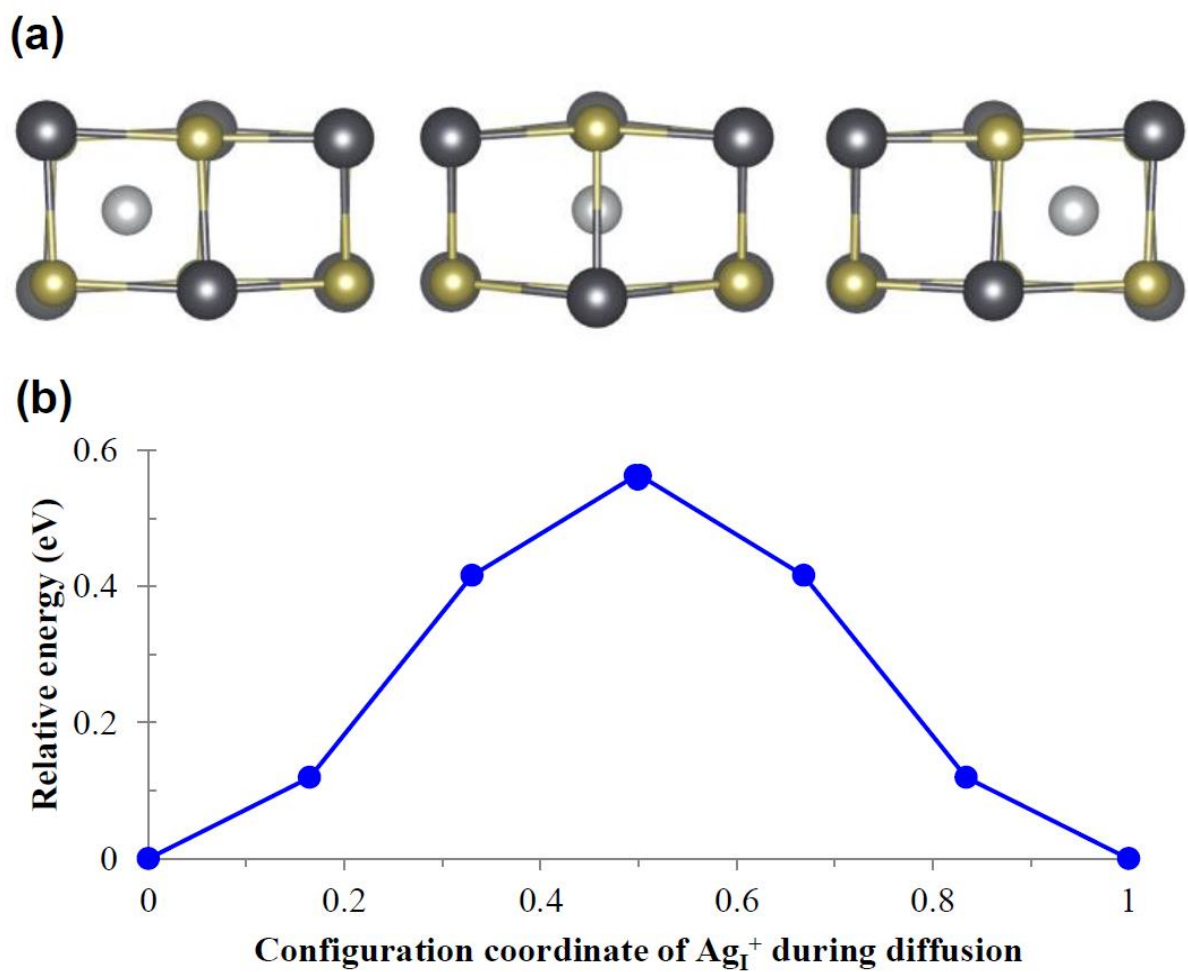

**FIGURE 3**

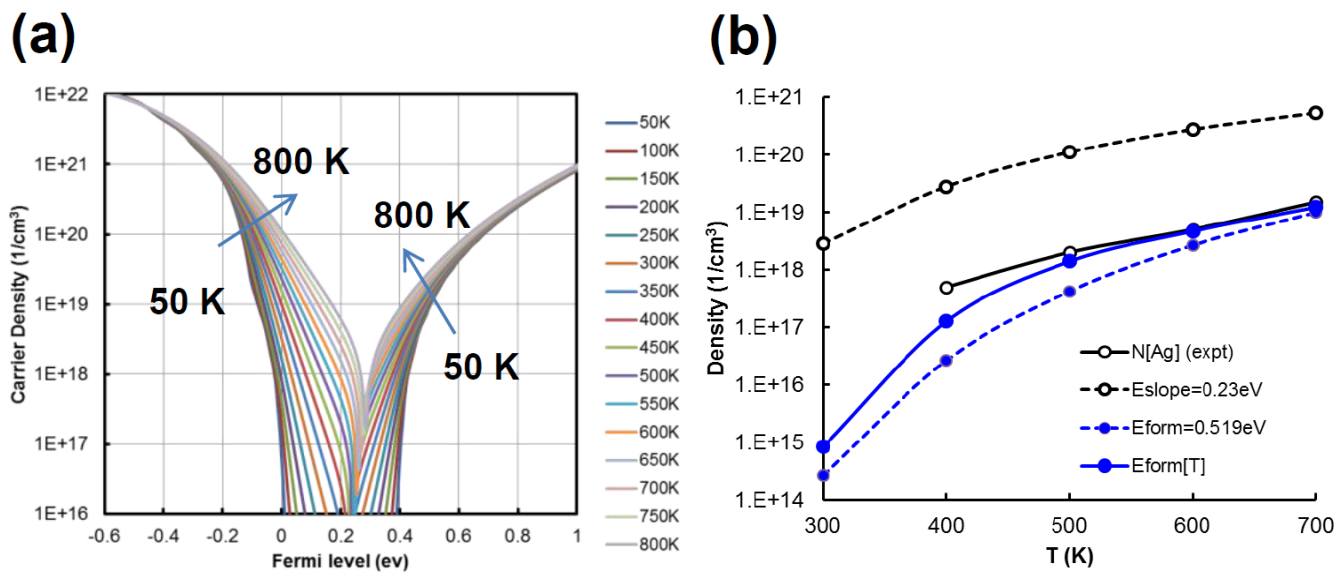

**FIGURE 4**